\documentclass{desyproc}

\usepackage{ulem}
\usepackage{soul}
\usepackage{color}

\newcommand{\GeV}{\mathinner{\mathrm{GeV}}}
\newcommand{\TeV}{\mathinner{\mathrm{TeV}}}

\def\l{\left}
\def\r{\right}

\def\bea{\begin{eqnarray}}
\def\eea{\end{eqnarray}}
\def\beq{\begin{equation}}
\def\eeq{\end{equation}}

\begin{document}
\title{Singlet portal extensions of the standard seesaw models to dark sector 
with local dark symmetry: \\ An alternative to the new minimal standard model}

\author{{\slshape Seungwon Baek, Pyungwon Ko\footnote{
Speaker. 9th Patras Workshop on Axions, WIMPs and WISPs}, Wan-Il Park}\\[1ex]
School of Physics, KIAS, Seoul 130-722, Korea}

\contribID{familyname\_firstname}

\desyproc{DESY-PROC-2013-XX}
\acronym{Patras 2013} 
\doi  

\maketitle

\begin{abstract}
Assuming dark matter is absolutely stable due to unbroken dark gauge symmetry 
and singlet operators are portals to the dark sector, we  present a simple extension 
of the standard seesaw model that can accommodate all the cosmological observations 
as well as terrestrial experiments available as of now, including  leptogenesis, 
extra dark radiation of $\sim 0.08$ (resulting in $N_{\rm eff} = 3.130$ the effective 
number of neutrino species),  Higgs inflation, small and large scale structure formation, 
and current relic density of  scalar DM ($X$).  
The Higgs signal strength is equal to one as in the SM for unbroken 
$U(1)_X$ case with a scalar dark matter,  but it could be less than one independent of 
decay channels if the dark matter is a dark sector  fermion or if $U(1)_X$ is spontaneously 
broken, because of a mixing with  a new neutral scalar boson in the models.   
\end{abstract}

\section{Introduction}

The standard model (SM) based on $SU(3)_C \times SU(2)_L \times U(1)_Y$ is 
confirmed at quantum level with high accuracy.
%
Still the SM has to be extended in order to accommodate the following observations:
($i$) neutrino masses and mixings, ($ii$) baryogenesis, ($iii$) nonbaryonic cold dark matter 
of the universe, and ($iv$) inflation and density perturbation. 

For the 1st and the 2nd items, the most economic and aesthetically attractive idea is 
to introduce singlet right-handed neutrinos and the seesaw mechanism, and  
leptogenesis for baryon number asymmetry.  For the 3rd item, there are many models
for cold dark matter, from axion to lightest superparticles to hidden sector DMs, to 
name a few.  For the 4th item the simplest inflation model without new inflaton fields 
would be $R^2$ inflation by Starobinsky~\cite{R2} and Higgs 
inflation~\cite{Bezrukov:2007ep}. 

In nonsupersymmetric dark matter models, one often assumes ad hoc $Z_2$ symmetry
in order to stabilize DM, without deeper understanding of its origin or asking if it is global 
or local discrete symmetry.  If we assume that global symmetry is not protected by 
quantum gravity effects, this $Z_2$ symmetry would be broken by $1/M_{\rm Planck}$ suppressed nonrenormalizable operators~\cite{Baek:2013qwa}. 
Then the electroweak scale DM can not live long enough to be dark matter candidate 
of the universe. The simplest way to guarantee the stability of EW scale DM is to assume 
the DM carries its own gauge charge which is absolutely conserved. 
Thus we are led to local dark symmetry and dark gauge force. 
This would be a very natural route for the DM model building, since the unsurpassed 
successful SM is also based on local gauge symmetry and its spontaneous breaking.

If weak scale DM carried nonzero SM gauge charges, it would be strongly constrained by 
direct detection cross section as well as electroweak precision observables and flavor
physics. Therefore we assume the DM is neutral under the SM gauge interaction, 
and making a hidden sector.  A hidden sector is generic in many models beyond the SM, 
including SUSY models or superstring theories. 
For example, high rank gauge group in the string theory would eventually break down
to $G_{\rm SM} \times G_{\rm hidden}$, where $G_{\rm hidden}$ is nothing but 
the dark gauge symmetry acting on hidden sector dark matter. 
If $G_{\rm hidden}$ is unbroken, 
DM particles would be absolutely stable, like the electron is absolutely stable due to 
electric charge conservation. If dark gauge coupling is strong and 
dark gauge interaction is confining like ordinary QCD, the DM would be the lightest 
composite hadrons in the hidden sector.  In this case it is possible to generate all 
the mass scales of the SM particles as well as the DM mass from dimensional 
transmutation in the hidden sector strong interaction~\cite{Hur:2011sv}. 
If dark gauge coupling is weak, we can employ the standard perturbation method 
to analyze the problems, which we adopt in the model described in this talk.

Another guiding principle is renormalizability of the model. The present authors found 
that one would get erroneous results if the effective Lagrangian approach is used  
for singlet fermion or vector DM with Higgs portal~\cite{Baek:2011aa,Baek:2012se}.  

Finally we generalize the notion of Higgs portal to the singlet portal,  assuming that 
the singlet operators in the standard seesaw model make portals to the dark sector. 
Note that there are only 3 singlet operators: $H^\dagger H, N_R$ and the kinetic mixing
between $U(1)_X$ and $U(1)_Y$ field strength tensors.

In this talk, I present a simple renormalizable model where the dark matter lives in a 
dark (hidden) sector with its own dark gauge charge along with dark gauge force. 
We mainly discuss the unbroken $U(1)_X$ dark gauge symmetry, and briefly mention
what happens if $U(1)_X$ is spontanesouly broken. This talk is based on 
Ref.~\cite{Baek:2013qwa}, to which we invite the readers for more detailed discussions
on the subjects described in this talk.

\section{Model}
As explained in Introduction, we assume that dark matter lives in a hidden sector, 
and it is stable due to unbroken local $U(1)_X$ dark gauge symmetry.  
All the SM fields are taken to be $U(1)_X$  singlets. Assuming that the RH neutrinos are 
portals to the hidden sector, we need both a scalar ($X$) 
and a Dirac fermion ($\psi$) with the same nonzero dark charge. 
Then the composite operator $\psi X^\dagger$ becomes a gauge singlet and thus can 
couple to the  RH neutrinos $N_{Ri}$'s  
\footnote{If we did not assume that the RH neutrinos are portals to the dark sector, 
we did not have to introduce both $\psi$ and $X$ in the dark sector. This case is discussed 
in brief in Sec.~3, Table~1.}.

With these assumptions, we can write the most general renormalizable Lagrangian as follows:   
\beq \label{Lagrangian}
\mathcal{L} = \mathcal{L}_{\rm SM} + \mathcal{L}_X + {\mathcal{L}_\psi} 
+  \mathcal{L}_{\rm portal} 
+ \mathcal{L}_{\rm inflation}
\eeq
where $\mathcal{L}_{\rm SM}$ is the standard model Lagrangian and  
\bea
\label{LX}
\mathcal{L}_X &=& {\l| \l( \partial_\mu + i g_X q_X \hat{B}'_\mu \r) X \r|^2} - \frac{1}{4} \hat{B}'_{\mu \nu} \hat{B}^{'\mu \nu} - m_X^2 X^\dag X - \frac{1}{4} 
\lambda_X \l( X^\dag X \r)^2
\\
\mathcal{L}_\psi &=& i \bar{\psi} \gamma^\mu \l( \partial_\mu + i g_X q_X \hat{B}'_\mu \r) \psi - m_\psi \bar{\psi} \psi 
\\
\label{portal}
\mathcal{L}_{\rm portal} &=& - \frac{1}{2} \sin \epsilon \hat{B}'_{\mu \nu} \hat{B}^{\mu \nu}
\nonumber 
- \frac{1}{2} \lambda_{HX} X^\dag X H^\dag H
\\
& - & \frac{1}{2} M_i \overline{N_{Ri}^C} N_{Ri} + \left[ Y_\nu^{ij} \overline{N_{Ri}} 
\ell_{Lj} H^\dag + \lambda^i \overline{N_{Ri}} \psi X^\dag + \textrm{H.c.} \right] 
\label{}
\\
\mathcal{L}_{\rm inflation} & = & [ \xi_H H^\dagger H + \xi_X X^\dagger X ] R
\eea
$g_X$, $q_X$, $\hat{B}'_\mu$ and $\hat{B}'_{\mu \nu}$ are the gauge coupling, 
$U(1)_X$ charge, the gauge field and the field strength tensor of the dark $U(1)_X$, 
and $R$ is the scalar curvature, respectively. 
$\hat{B}_{\mu \nu}$ is the gauge field strength of the SM $U(1)_Y$.
We assume $m_X^2 > 0, \quad \lambda_X > 0, \quad \lambda_{HX} > 0$,  so that the local 
$U(1)_X$ remains unbroken and the scalar potential is bounded from below at tree level.

This model has only 3 more fields compared to the standard seesaw models, and is 
based on local gauge principle for  absolutely stable DM rather than ad hoc $Z_2$ 
symmetry~\cite{Davoudiasl:2004be}.   Therefore, our model could
be considered as an alternative to the so-called new minimal SM~\cite{Davoudiasl:2004be}.

\section{Implications on particle physics and cosmology}

Our model is simple enough, but has sufficiently rich structures, and it can accommodate
various observations from cosmology and astrophysics with definite predictions for 
Higgs physics:
\begin{itemize}
\item Dark scalar $X$ can improve the stability of the electroweak vacuum up to Planck scale, unlike the SM for $m_h \sim 125 \GeV$ and $m_t = 173.2 \GeV$ and $\alpha_s = 0.118$, 
if $\lambda_X > 0$ and $\lambda_{HX} \gtrsim 0.2$.  
\item Perturbativity of quartic couplings for scalar fields $H$ and $X$ up to Planck scale puts theoretical constraints on $\lambda_X$ and $\lambda_{HX}$ such that $\lambda_X \lesssim 0.2$ and $\lambda_{HX} \lesssim 0.6$.
\item Massless dark photon mediates long range between dark matter, and can solve the small scale problem of DM subhalo while satisfying constraints from inner structure and kinematics of dark matter halos,
if $g_X \lesssim 2.5 \times 10^{-2} \left( m_X/300 \GeV \right)^{3/4}$.
\item If dark fermion $\psi$ were lighter than $X$ and became DM, then its thermal 
relic density would be too large since it can annihilate only into a pair of dark photon 
($\sigma_{\rm ann} v \propto g_X^4$). On the contrary the dark scalar $X$ can be diluted
efficiently even if $g_X$ is very small, since there is a Higgs portal term which makes 
$X X^\dagger \leftrightarrow$ (SM particles).
\item Direct detection experiments such as XENON100 and CDMS put strong bounds  
on the combination of the gauge kinetic mixing $10^{-12} \lesssim \epsilon g_X \lesssim 10^{-5}$ for $6 \GeV \lesssim m_X \lesssim 1 \TeV$ when the upper bound on $g_X$ is used.
\item Massless dark photon in our model would contribute to the number of effective neutrinos 
which can be measured accurately by Planck satellite and others. We find that dark photon 
contributes to dark radiation by $\sim 0.08$, which is in agreement with the recent 
Planck data, 
$\Delta N_{\rm eff} = 3.30 \pm 0.27$ at $68\%$ CL.
\item The decay of right-handed(RH) neutrinos generate both matter and dark matter 
thanks to see-saw mechanism.
However the asymmetric component of dark matter disappears as the heavy dark fermion 
$\psi$ decays eventually, that 
can also generate visible sector lepton number asymmetry  large enough 
to match the observation. 
\item Higgs inflation can work in our model since the gauge singlet scalar coupled to SM Higgs field cures the instability of potential in Higgs-singlet system.
Inflation along the SM Higgs direction does not pose any new constraint on the model parameters.  
\item In case $U(1)_X$ is unbroken, the Higgs signal strength should be equal to ``1'', 
independent of production and decay channels.  If we consider other variations of the
model with broken $U(1)_X$ or only dark scalar or dark fermion, the number of Higgs-like
scalar bosons can be more than one, with universally reduced Higgs signal strength    
(see Table~1). 
\end{itemize}

In conclusion, we presented a simple extension of the standard seesaw model 
where dark matter physics is constructed with local dark gauge symmetry.  
The predictions in various cases are summarized in Table~1. 

\begin{table}[htdp]
\begin{center}
\begin{tabular}{|c|c|c|c|c|c|}
\hline
Dark sector fields & $U(1)_X$ & Messenger & DM & Extra DR & $\mu_i$  
\\   \hline  
$\hat{B}'_\mu , X , \psi$ & Unbroken & $H^\dagger H  , \hat{B}'_{\mu\nu} \hat{B}^{\mu\nu} , N_R$ 
& $X$ & $\sim 0.08$ & $1~(i=1)$
\\
$\hat{B}'_\mu , X$ & Unbroken & $H^\dagger H , \hat{B}'_{\mu\nu} \hat{B}^{\mu\nu}$ 
& $X$ & $\sim 0.08$ & $1 ~(i=1)$
\\
$\hat{B}'_\mu , \psi$ & Unbroken & $H^\dagger H  , \hat{B}'_{\mu\nu} \hat{B}^{\mu\nu} ,  S$ 
& $\psi$ & $\sim 0.08$ & $< 1~(i=1,2)$
\\
\hline 
$\hat{B}'_\mu , X , \psi , \phi$ & Broken & 
$H^\dagger H  , \hat{B}'_{\mu\nu} \hat{B}^{\mu\nu} , N_R$ 
& $X$ or $\psi$ & $\sim 0$ & $< 1~(i=1,2)$
\\
$\hat{B}'_\mu , X , \phi$ & Broken 
& $H^\dagger H , \hat{B}'_{\mu\nu} \hat{B}^{\mu\nu}$ & $X$ & $\sim 0$ & $< 1 ~(i=1,2)$
\\
$\hat{B}'_\mu , \psi$ & Broken 
& $H^\dagger H  , \hat{B}'_{\mu\nu} \hat{B}^{\mu\nu} ,  S$ 
& $\psi$ & $\sim 0$ & $~~< 1~(i=1,2,3)$
\\   \hline
\end{tabular}
\end{center}
\caption{Dark fields in the hidden sector, messengers, dark matter (DM), the 
amount of dark radiation (DR), and the signal strength(s) of the $i$ scalar boson(s) 
($\mu_i$)  for unbroken or spontaneously broken (by $\langle \phi \rangle \neq 0$) 
$U(1)_X$ models considered 
in this work.  The number of Higgs-like neutral scalar bosons could be 1,2 or 3, 
depending on the  scenarios. }
\label{default}
\end{table}%



\section{Acknowledgments}
This work was supported in part by 
National Research Foundation of Korea (NRF) Research Grant 2012R1A2A1A01006053, 
and by SRC program of NRF funded by MEST (20120001176) through Korea Neutrino 
Research Center at Seoul National University (PK).  

\begin{footnotesize}

\end{footnotesize}


\end{document}